\documentclass{article}[12pt]
\usepackage{amsfonts,amsmath}

\newtheorem{theorem}{Theorem}

\def\R{{\mathbb R}}
\def\C{{\mathbb C}}

\begin{document}

\title{The Moutard transformation and two-dimensional
multi-point delta-type potentials}
\author{R.G. Novikov \thanks{CNRS (UMR 7641), 
Centre de Math\'ematiques Appliqu\'ees, Ecole
Polytechnique, 91128 Palaiseau, France, and Faculty of Control and Applied Mathematics, MIPT, 141700 Dolgoprudny , Russia; 
e-mail: novikov@cmap.polytechnique.fr}
\and 
I.A. Taimanov \thanks{Sobolev Institute of Mathematics, 630090 Novosibirsk, Russia; e-mail:
taimanov@math.nsc.ru}}
\date{}
\maketitle

Let $H$ be a two-dimensional Schr\"odinger operator 
$H = -\Delta + U = -4 \bar{\partial}\partial + U$,
where 
$\partial = \frac{1}{2}\left(\frac{\partial}{\partial x} - 
i \frac{\partial}{\partial y}\right),
x,y \in \R$,
and let $\omega$ be a formal solution to the equation
\begin{equation}
\label{10}
H \omega = 0.
\end{equation}
The Moutard transformation corresponds to $H$ and $\omega$ the operator
\begin{equation}
\label{1}
\widetilde{H} = -4 \bar{\partial} \partial + \widetilde{U} = 
-4 \bar{\partial}\partial + U - 8\bar{\partial}\partial\log\omega
\end{equation}
such that for every $\varphi$ meeting the equation $H\varphi=0$ 
a function $\theta$ satisfying the system
\begin{equation}
\label{2}
(\omega\theta)_z = -i\omega^2 \left(\frac{\varphi}{\omega}\right)_z, \ \ \
(\omega\theta)_{\bar{z}} = i\omega^2 \left(\frac{\varphi}{\omega}\right)_{\bar{z}},
\end{equation}
satisfies $\widetilde{H}\theta = 0$. 
The function $\theta$ is defined modulo $\frac{1}{\omega}$ 
due to the integration constant in the right-hand sides of (\ref{2}).

Recently the Moutard transformation which originates in the surface theory
was used for constructing special types of two-dimensional potentials
and blowing up solutions of the Novikov--Veselov equation \cite{TT,TT1}.

In difference with \cite{TT} which concerns with regular potentials 
in the present note we deal with multi-point delta-type potentials. 
We consider also the Faddeev eigenfunctions  \cite{Faddeev1965}
of the corresponding operators $H$
on the zero energy level. 
These eigenfunctions are defined by 
conditions
$$
H\psi = 0, \ \ \ 
\psi(z,\bar{z},\lambda) = 
e^{\lambda z}(1 + o(1)) \ \ \ \mbox{as $z \to \infty$},\ 
\lambda \in\C \setminus\{0\}.
$$
In addition,
$$
\psi=e^{\lambda z}\biggl(1+\frac{a(\lambda,\bar\lambda)}{z}+
e^{\bar\lambda\bar z-\lambda z}\frac{b(\lambda,\bar\lambda)}{\bar{z}}+
o\biggl(\frac{1}{|z|}\biggr)\biggr)\ \ {\rm as}\ \ z\to\infty,
$$
where $a,b$ are the Faddeev generalized "scattering" data on the zero energy
level.

\begin{theorem}
A formal application of the Moutard transformation to the zero potential 
$U=0$ by using a polynomial in $z$ function $\omega = P(z) = 
\prod_{k=1}^N (z-z_k)$ leads to the milti-point delta-type potential
$$
\widetilde{U}(z) = - 8\pi \sum_{k=1}^N\delta(z-z_k).
$$
For this potential the Faddeev eigenfunctions on the zero energy level take the form
\begin{equation}
\label{3}
\psi = e^{\lambda z}\left(1+ 
\frac{2}{P}\sum_{k  = 1}^N \frac{(-1)^k P^{(k)}(z)}{\lambda^k}\right),
\end{equation} 
where $P^{(k)}(z) = \partial^k P(z)$.
In addition, for these eigenfunctions
$a = -2N/\lambda$ and  $b\equiv 0$.
\end{theorem}

The proof of this theorem is based on solving system (\ref{2}) with respect to
$\psi=\theta$ for $\omega=P(z)$ and $\varphi=ie^{\lambda z}$, and on 
straightforward computations. 
However we need
to clarify the meaning of Schr\"odinger operators with such potentials.

Actually in this case we consider the Moutard system (\ref{2})
with $\omega = P(z)$ as the appropriate regularization of the
Schr\"odinger equation $\widetilde{H}\theta = 0$ with the potential 
$\widetilde{U}$ of Theorem 1.
In addition, for $N=1$, the Schr\"odinger equation 
$(-\Delta + \widetilde{U})\psi =0$ with $\widetilde{U}$ and 
$\psi$ from Theorem 1 is formally fulfilled under the following conventions:
$$
\bar\partial\biggl(e^{\lambda z}\biggl(\frac{1}{z}\biggr)^2\biggr)=
e^{\lambda z}\frac{2}{z}\bar\partial\biggl(\frac{1}{z}\biggr)=
\frac{2\pi e^{\lambda z}\delta(z)}{z}.
$$

We remark that the functions $\psi$ of (\ref{3}) essentially differ from 
the Faddeev eigenfunctions found in \cite{GNR,GNR1} 
for the Schr\"odinger operators with multi-point delta-type potentials. 
The reason is that in \cite{GNR,GNR1} the operator with such a potential is 
replaced by its regularization going back to \cite{BF}, whereas in the
present note we work formally with the original potentials 
considering the regularization, of the equation $\widetilde{H}\theta =0$,
given by the Moutard system (\ref{2}). 

In addition, in view of the property $b\equiv 0$ for $\psi$ of (\ref{3}) the
potentials of theorem 1 may be considered  as "reflectionless" in the sense
of the Faddeev generalized "scattering" data $a,b$. In this sense the
functions $\psi$ of (\ref{3}) are similar to the Faddeev eigenfunctions found
in \cite{TT3} for some regular potentials.

In \cite{TT,HLL} the Moutard transformation is extended to 
a transformation of solutions of the Novikov--Veselov equation \cite{VN}
\begin{equation}
\label{5}
U_t = \partial^3U + \bar{\partial}^3U + 3\partial(UV) + 
3 \bar{\partial}(U\bar{V}) = 0, \ \ \ \ -4\bar{\partial}V = \partial U.
\end{equation}
This equation has the Manakov form $H_t = HA+BH$ 
where $A$ and $B$ are differential operators. 
If $U$ satisfies (\ref{5}) and $\omega$ meets (\ref{10}) and the equation 
\begin{equation}
\label{6}
(\partial_t +A)\omega =0, 
\end{equation}
then the extended Moutard transformation 
of $U$ has the same form (\ref{1}) 
and gives a new solution of (\ref{5}). 
For the zero potential $U=V=0$, we have 
$A=\partial^3 + \bar{\partial}^3$ and  
$\omega(z,t) = P(z,t) = 
\prod_{k=1}^N  (z-z_k(t))$ satisfies (\ref{6}) if and only if
$$
\frac{\partial P}{\partial t} = \frac{\partial^3 P}{\partial z^3}.
$$
The latter equation describes an algebraic dynamics of the zeroes of $P(z,t)$
(such a dynamics for another reason was considered in \cite{TT}).
A formal application of the extended Moutard transformation leads to 
the potential 
$$
\widetilde{U}(z,t) = -8\pi \sum_{k=1}^N \delta(z-z_k(t))
$$
which apparently may be considered as a formal solution to (\ref{5}).

{\bf Acknowledgement.} 
This work was done during the visit of one of the
authors (I.A.T.)  to Centre de Math\'ematique Appliqu\'ees of Ecole
Polytechnique in July, 2013.
The work was partially supported by TFP No 14.A18.21.0866 of 
Ministry of Education and Sciences of
 Russian Federation  (R.G.N.) and by 
the program ``Fundamental problems of nonlinear dynamics'' of 
the Presidium of RAS and 
the grant 1431/GF of Ministry of Education and Science of 
Republic of Kazakhstan (I.A.T.).


\begin{thebibliography}{MMM}


\bibitem{TT}
Taimanov, I.A., and Tsarev, S.P.: 
Two-dimensional rational solitons and their blow-up via 
the Moutard transformation. 
Theoret. and Math. Phys. {\bf 157} (2008), 1525-1541. 

\bibitem{TT1}
Taimanov, I.A., and Tsarev, S.P.:
Blowing up solutions of the Novikov-Veselov equation. 
Doklady Math. {\bf 77} (2008), 467-468.

\bibitem{Faddeev1965}
Faddeev, L.D.:
Growing solutions of the Schr\"odinger equation.
Soviet Phys. Dokl. {\bf 10} (1965), 1033--1035.

\bibitem{GNR}
Grinevich, P.G., and Novikov, R.G.:
Faddeev eigenfunctions for point potentials in two dimensions.
Physics Letters A {\bf 376} (2012), 1102--1106.

\bibitem{GNR1}
Grinevich, P.G., and Novikov, R.G.:
Faddeev eigenfunctions for multipoint potentials.
arXiv:1211.0292.

\bibitem{BF}
Berezin, F.A., and Faddeev, L.D.:
Remark on Schr\"odinger equation with
singular potential. 
Soviet Mathematics {\bf 2} (1961), 372--375.

\bibitem{VN}
Novikov, S.P., and Veselov, A.P.:
Finite-zone, two-dimensional potential Schr\"odinger operators.
Explicit formulas and evolution equations.
Soviet Math. Dokl. {\bf 30} (1984), 588--591.

\bibitem{HLL}
Hu Heng-Chun, Lou Sen-Yue, and Liu Qing-Ping:
Darboux transformation and variable separation approach:
the Nizhnik-Novikov-Veselov equation.
Chinese Phys. Lett. {\bf 20} (2003), 1413--1415.

\bibitem{TT3}
Taimanov, I.A., and Tsarev, S.P.:
Faddeev eigenfunctions for two-dimensional Schr\"odinger 
operators via the Moutard transformation.
arXiv:1208.4556.



\end{thebibliography}
\end{document}